\documentclass[prb,aps,amsmath,amssymb,amsfonts,floatfix,groupedaddress,showpacs,twocolumn]{revtex4}
\usepackage{graphicx}
\usepackage[dvips]{color}
\usepackage{dcolumn}
\def\srt3{$\sqrt{3}$$\times$$\sqrt{3}R30\,\mathring{}$}
\def\rt3{3$\times$3}

\def\t2g{$t_{2g}$}

\begin{document}

\title{Realistic modeling of the electronic structure and the effect of correlations 
for Sn/Si(111) and Sn/Ge(111) surfaces}
\author{Sergej Schuwalow}
\affiliation{I. Institut f{\"u}r Theoretische Physik, 
Universit{\"a}t Hamburg,
D-20355 Hamburg, Germany}
\author{Daniel Grieger}
\affiliation{I. Institut f{\"u}r Theoretische Physik, 
Universit{\"a}t Hamburg,
D-20355 Hamburg, Germany}
\author{Frank Lechermann}
\affiliation{I. Institut f{\"u}r Theoretische Physik, 
Universit{\"a}t Hamburg,
D-20355 Hamburg, Germany}

\begin{abstract}
The correlated electronic structure of the submonolayer surface systems Sn/Si(111) 
and Sn/Ge(111) is investigated by density-functional theory (DFT) and its 
combination with explicit many-body methods. Namely, the dynamical mean-field 
theory and the slave-boson mean-field theory are utilized for the study of the 
intriguing interplay between structure, bonding and electronic correlation. In this
respect, explicit low-energy one- and four($sp^2$-like)-band models are derived 
using maximally-localized Wannier(-like) functions. In view of the possible 
low-dimensional magnetism in the Sn submonolayers we compare different types of 
magnetic orders and indeed find a 120$^\circ$ antiferromagnetic ordering to be 
stable in the ground state. With single-site methods and cellular-cluster 
extensions the influence of a finite Hubbard $U$ on the surface states in a planar 
and a reconstructed structural geometry is furthermore elaborated.

\end{abstract}

\maketitle

\section{Introduction}
Strong electronic correlations lead to many of the most interesting
phenomena in modern condensed matter physics, such as superconductivity, 
the metal-insulator transition or local moment magnetism. It is well known
that the effect of these many-body effects depend heavily on the 
interplay between structure, bonding, and the degree of order within the
effective dimensionality of the given problem. While the study of 
realistic bulk systems in this context has been rather extensive in recent
years, those investigations may only provide insight into the effect of
dimensionality via reasonably justified quasi-dimensions due to large 
crystalline anisotropies. In this respect the study of surface systems has
the advantage that by an intelligible tuning of the coupling between the
adsorbate and the substrate, quasi-lowdimensionality can be created more
efficiently. Besides, there are many surface-sensitive experimental probe
techniques in order to reveal the electronic structure more directly than
in the bulk. Hence the research on specific surface materials problems 
may add substantially to the understanding of the general problem of strong 
quantum correlations.

There has been an increased interest in adsorbate systems involving
semiconducting substrates since more than a decade~\cite{car96,wei97,zha10}. These
often exhibit dangling-bond-derived surface states with rather small bandwidths. 
While Mott criticality is hardly seen on the free surfaces because of structural 
reconstructions, certain adsorbate atoms forming (sub)monolayers stabilize 
hybridized narrow-band surface states down to low temperatures. The most prominent 
of this kind are the so-called $\alpha$-phase surfaces~\cite{car96} where the 
canonical structure is described by a \srt3 triangular array of adsorbate atoms 
within an 1/3 monolayer coverage on a (111) semiconductor substrate, resulting in a 
half-filled surface band. Due to the rather large interatomic distances between the 
adsorbate atoms ($\sim$7\AA) the latter becomes indeed rather narrow.  
Mott-insulating phases in such systems are believed to be realized in e.g., 
K/Si(111)~\cite{wei97}, whereas a transition to a correlation-driven 
charge-density-wave phase takes place in Pb/Ge(111)~\cite{car96}.

In this respect, the $\alpha$-phase systems Sn/Si(111) and Sn/Ge(111) are of central
interest due to the unusual properties they exhibit at low temperatures. It has 
been recently verified by Modesti {\sl et al.}~\cite{mod07} that the planar 
$\alpha$-Sn/Si(111) surface exhibits a metal-insulator transition below 60 K. 
No structural reconstruction appears to accompany this transition, since it was 
observed~\cite{mor02} that the \srt3 periodicity is stable at least down to around 
6 K. The question of magnetic ordering within the insulating regime, i.e., the 
formation of an antiferromagnetic (AFM) Mott-insulating state, has been raised 
because of band foldings revealed by low $T$ photoemission 
experiments~\cite{uhr00,cha01,mod07} leading to an \rt3 periodicity. 
In contrast, the structurally and electronically very similar $\alpha$-Sn/Ge(111) 
surface shows vastly different behaviour. There a transition from the \srt3 
phase at room temperature to an \rt3 symmetry below 200K takes place.
Two competing ground-state configurations of the Sn atoms seem to exist in the 
latter temperature range~\cite{avi99}, with the so called 2D-1U 
(i.e., two Sn atoms down, one Sn atom up with respect to the planar triangular
structure) state in favor of 6 meV/adatom against the competing 1D-2U 
configuration~\cite{pul06}. The \srt3 periodicity observed at room temperature is
understood as a result of the rapid fluctuations of the system between these two 
states~\cite{avi99}. It is still a matter of debate whether or not the system 
displays a surface Mott transition similar to the Sn/Si(111) system. The finding of 
a Mott insulating phase in $\alpha$-Sn/Ge(111) below 20K by 
Cortes {\sl et al.}~\cite{cor06} has apparently not been confirmed by other 
groups~\cite{col08,mor08}.

The electronic structure of the $\alpha$ surface phases poses an interesting 
problem in the context of strong electron correlation. Early
phase-diagram studies of such adlayer structures within the Hartree-Fock 
approximation showed the possibility for various orderings~\cite{san99}. Since the 
narrow-band surface state is mainly derived from a hybridization of the Sn($5p_z$) 
state with the underlying Si states, one deals with Coulomb correlations in an 
effective $5p$ system (assuming a local Hubbard-like interaction). 
Calculations based on standard density-functional theory (DFT) 
show~\cite{pro00,flo01} in the case of \srt3 Sn/Si(111) indeed an isolated 
half-filled surface band of width $W$$\sim$0.3 eV. Additionally, the named \rt3 
reconstruction in Sn/Ge(111) is verified within 
DFT~\cite{avi99,gir00,per01,bal02,pul06,gor09}.
Correlation effects beyond the local density approximation (LDA) and the 
generalized gradient approximation (GGA) have been investigated by Profeta and 
Tosatti~\cite{pro07} within the LDA+U method~\cite{ani_ldau}, establishing a 
Hubbard $U$ for the single Sn($5p_z$) orbital. In this 
scheme, Mott-insulating Sn/Si(111) (with assuming ferro-/ferrimagnetic order) is 
reached for $U$$\simeq$2 eV, but the authors find a $U$$\simeq$4 eV 
from constrained calculations more appropriate, also to cope with a gap size of 
order $\sim$0.3 eV. 
This rather large value for the local Coulomb interaction differs from 
$U_{\rm eff}$$\simeq$1.15 eV obtained in elder constrained LDA calculations 
designed for a minimal model describing the surface band~\cite{flo01}. Concerning
$\alpha$-Sn/Si(111) there are speculations~\cite{san99,mod07,pro07} about a 
realization of the quasi-twodimensional (2D) correlated triangular lattice problem,
including the possiblity of antiferromagnetism or spin-liquid physics and 
$d$-wave superconductivity. 

In this work we want to investigate the importance of electronic correlations 
and their interplay with the structural data in the $\alpha$-Sn/(Si,Ge)(111) phases.
By means of a combination of realistic DFT band-structure schemes with advanced 
many-body techniques, the aim is to clarify the differences stemming from the 
(Si,Ge)(111) substrates and to reveal to which extent Coulomb correlations can give 
rise to the observed and perhaps still-to-be obeserved phenomena. Albeit some
relevant work has already been performed in this direction, namely 
Ref.~\onlinecite{san99,flo01,pro07}, there remain still many open questions. 
For instance,
the minimal Hubbard model derived by Flores {\sl et al.}~\cite{flo01} has not 
truly been numerically treated and the arguments given concerning the influence of 
structural reconstructions on a possible Mott-criticality lack a local picture of
the involved orbitals. The LDA+U method utilized in Ref.~\onlinecite{pro07} 
is designed for long-range-ordered insulating states and neglects as a static 
technique quantum fluctuations (even in the Mott state). Hence correlations in the
metallic regime are usually described incorrectly and magnetic tendencies are often
overestimated. Moreover, the experimental data is far from being conclusive, e.g.,
concerning the appearance of local moment physics and eventual magnetic ordering. 

\section{Theoretical Approach}

Realistic theoretical schemes for correlated condensed matter, combining 
traditional band-structure approaches with explicit many-body techniques 
have been quite successful in the last decade in dealing with various problems 
in strongly correlated physics. The most prominent of such schemes is the 
combination of DFT with the dynamical mean-field theory (DMFT), the so-called 
LDA+DMFT~\cite{ani97,lic98} framework.
While in standard DFT correlation effects are only taken into account in an 
averaged way by making reference to a homogeneous(-like) electron gas, the named
combined approach allows for explicit many-body effects on an operator level by 
still keeping important band-structure details from LDA. The DMFT 
technique~\cite{geo92,met89} is able to incorporate
all onsite quantum fluctuations and thereby may describe quasiparticle (QP) as well 
as atomic excitations (i.e., Hubbard bands) on an equal footing. 
A alternative combined approach is given by interfacing the Gutzwiller- or the 
slave-boson mean-field technique with DFT~\cite{bue03,den08,lec09}. In this more 
simplified treatment the QP lifetime remains infinite, thus omitting the full 
frequency dependence of the self-energy. Hence only low-energy features may be 
addressed in the spectral function and hence high-energy Hubbard bands are not
accessible. However importantly, the local atomic multiplets are still present with
an effective static character in the generalized theory~\cite{bue98,lec07,fab07}.

Due to the subtle dependencies of the $\alpha$-Sn/(Si,Ge)(111) surface 
electronic structure on the structural facts, care must be taken in its 
proper determination. We used an implementation~\cite{mbpp_code} of the 
highly-accurate mixed-basis pseudopotential (MBPP) technique~\cite{lou79} for this 
task. This DFT band-structure code employs normconserving 
pseudopotentials~\cite{van85} and an efficient combined basis consisting of plane 
waves and a few localized orbitals. Since the magnetism of these surface systems is 
very subtle, we additonally performed computations within the 
projector-augmented-wave (PAW) method~\cite{blo94} for the specific study of 
magnetic ordering. Thereby one is able to lift the possible limitations due to the 
use of pseudo-crystal wave functions. Two implementations of the PAW formalism, 
namely the CP-PAW~\cite{blo94} code and the Vienna Ab-initio Simulation Package 
(VASP)~\cite{kre94} code, were applied, which also allow for the investigation
of noncollinear spin orderings. The actual calculations were performed by employing a
slab geometry where care was taken in using a well converged lateral k-point mesh
(up to a 25$\times$25 grid for the magnetic structures).

We combined the DFT approach with two many-body techniques, namely the DMFT 
with a Hirsch-Fye quantum Monte-Carlo impurity solver~\cite{hir86} and the
rotationally invariant slave boson (RISB) method~\cite{li89,lec07} in saddle-point
approximation. Note that in the single-site problem the RISB technique may also 
be understood as a quasiparticle impurity solution to DMFT. The employed
mean-field version of the RISB method is in its character and nature of 
approximation very similar to the state-of-the-art Gutzwiller 
technique~\cite{bue98,bue07,fab07}. 

For the actual interfacing of DFT with the many-body approaches a suitable
downfolding procedure of the relevant problem to a local basis (the correlated
subspace) is needed 
(see e.g., Ref.~\onlinecite{lec06,kotliar_review,ani05}). The choice is in most 
cases a matter of convenience, since many physically sound frameworks that 
provide a representation of the band-structure within a minimal local 
Wannier(-like) basis are applicable. 
The coherent connection of the crystal problem to such a tailored basis is the key 
point, rather than the pecularities of the (restricted) projected local viewpoint
itself. It is important to realize that the extended (crystal, surface or chain) 
problem, although composed of atoms, is {\sl not} an atomic problem. Of course, in a 
subsequent many-body treatment the minimal interacting hamiltonian has to be 
adjusted to the characteristics of the chosen local basis. Up to now, the results 
of such many-body approaches are not exceedingly sensitive to the very details of 
the selected correlated subspace, once the number and character of orbitals is 
agreed on. In that sense, the here utilized maximally-localized Wannier (MLWF) 
scheme~\cite{mar97,sou01} thus provides a reliable local basis for the materials 
under consideration.

\section{DFT investigation and MLWF description}

\subsection{Single-site Sn unit cell\label{sec:single}}

\begin{figure}[b]
   \includegraphics[width=3.25in]{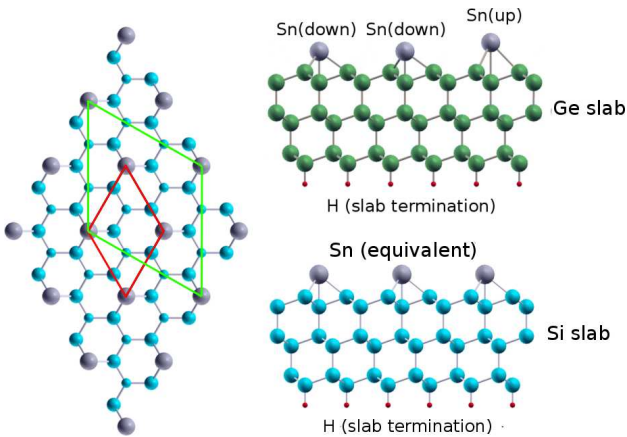}
  \caption{(Color online) 
  Left: Top view onto the employed \srt3 (red) and \rt3 (green) 
  surface unit cells . 
  Right: Side view of the slab geometry for Sn/Si(111) surface (bottom) and the 
  distorted 1U-2D Sn/Ge(111) surface (top).}
  \label{structure}
  \end{figure}
In order to investigate the electronic structure of the 1/3 monolayer of tin atoms
on the semiconductor \srt3 surfaces we first utilized slab geometries incorporating
a single Sn atom within the full unit cell. Thus the surfaces are modeled by a 
supercell with three bilayers of (Si,Ge) separated by sufficiently large vacuum 
regions. Note that in this section we model the Sn submonolayer in each case
as flat, although a distortion in $z$ direction is observed for true Sn/Ge(111)
(see section \ref{sec:cluster}). The Sn atoms are placed in the $T_{4}$ sites of 
the surface while the bottom of the slab is saturated with hydrogen atoms 
(see Fig.~\ref{structure}). Structural relaxations by minimizing the atomic forces 
have been performed with fixed position of the lowest substrate layer. The bulk 
chosen lattice constants and relaxed Sn-Sn nearest-neighbor distances are (5.43, 
6.65)\AA{} and (5.65, 6.93)\AA{} for Sn/Si(111) and Sn/Ge(111), respectively.
\begin{figure}[b]
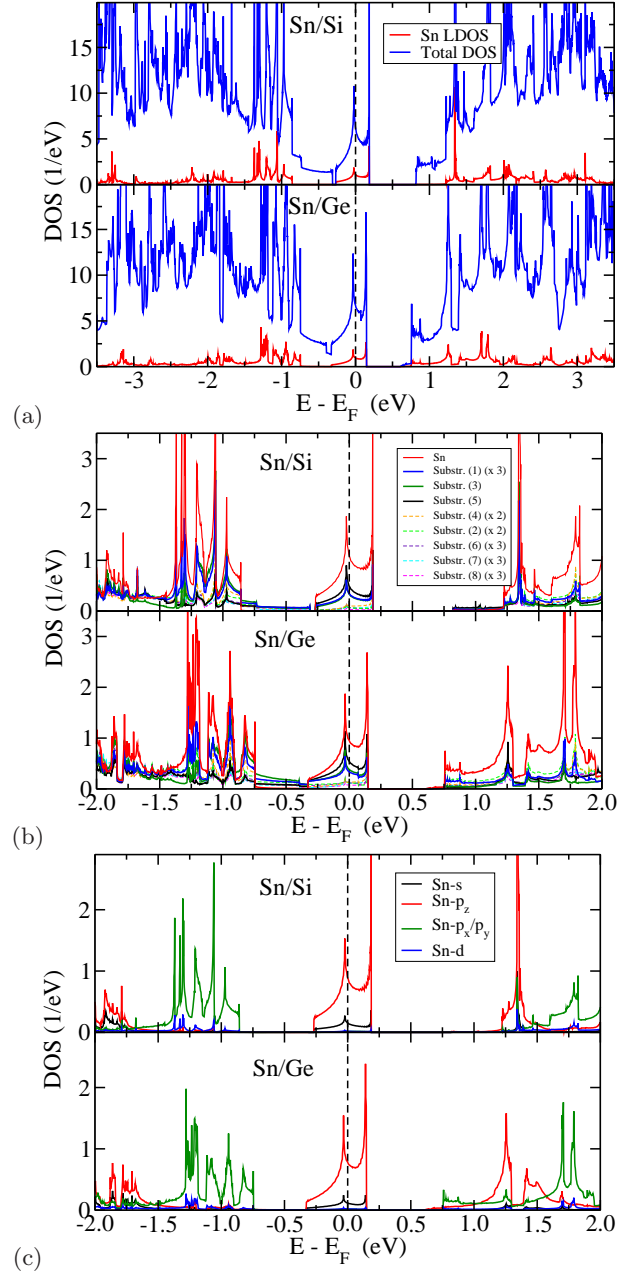

(a)\includegraphics*[width=3.00in]{Relative_DOS.eps}
(b)\includegraphics*[width=3.00in]{DOS_Plot.eps}
(c)\includegraphics*[width=3.00in]{Sn_DOS.eps}
\caption{(Color online) GGA-PBE DOS for Sn/(Si,Ge)(111) in the \srt3 structure. 
(a) Total DOS together with local Sn DOS and (b) shown via contributions from 
different atoms within the supercell (for the labeling of the substrate atoms see 
Fig.~\ref{wann-single}). (c) Angular-momentum resolved plot of the Sn DOS. 
Note that in the latter plot the  $5p_{x}$ and $5p_{y}$ curves lay on top of each 
other.}
\label{surfDOS}
\end{figure} 
An energy cutoff of 16 Ryd for the plane-wave part of the mixed-basis set was used 
for all calculations. Localized functions were introduced for Si($3s$,$3p$),
Ge($3d$,$4s$,$4p$) as well as Sn($4d$,$5s$,$5p$). Note that hence the semicore $d$ 
electrons of Ge and Sn are treated as valence in our computations and relativistic 
effects are included via the scalar-relativistic normconserving pseudpotentials. 
Since usually more appropriate for surface studies than LDA, we employed the GGA in 
the form of the PBE functional~\cite{per96} to the exchange-correlation term in
the density functional.

The density of states (DOS) shown in Fig.~\ref{surfDOS} displays for both surface 
systems a prominent structure close to the Fermi level of width $W$$\sim$ 0.4 eV 
(slightly larger for Sn/Ge(111)). Two subpeaks are visible, one nearby the
Fermi energy and the other at the upper energy edge in the unoccupied region.
The major contributions to this low energy part stems from the orbitals of the Sn 
adatom hybridizing with the orbitals of neighbouring substrate atoms of the first 
and second layer (see Fig.~\ref{surfDOS}b). It becomes further obvious from the 
angular-momentum resolved DOS in Fig.~\ref{surfDOS}c that concerning the tin part 
the Sn($5p_z$) orbital is dominantly responsible for the low-energy weight. Minor 
contribution to this is also added by the Sn($5s$) orbital. 

\begin{figure}[t]
 \includegraphics[width=3.25in]{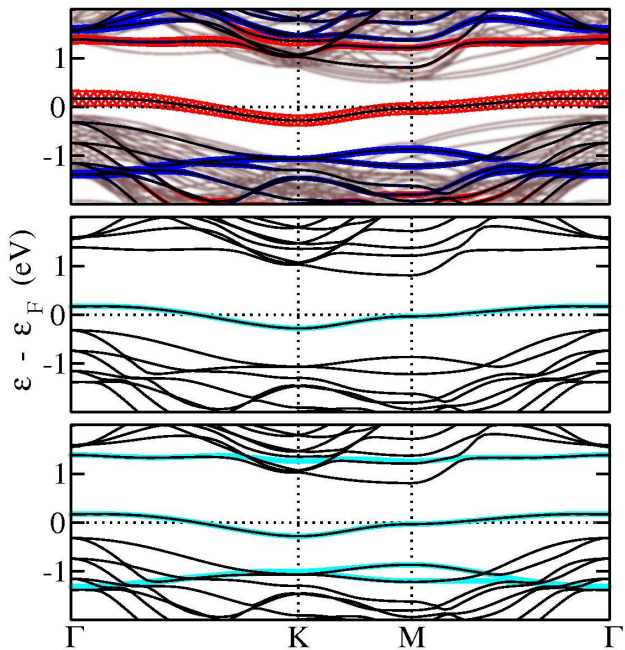}
\caption{(Color online) Top: Surface band structure for the Sn/Si(111)-\srt3 system,
with fatbands for the Sn($5p$) orbitals (blue/grey: $p_{x}$,$p_y$; 
red/lightgrey: $p_z$). The bulk Si band structure is shown in the background. 
Middle: Derived Wannier band (cyan/lightgrey) for the 1-band model. Bottom: 
Derived Wannier-like bands (cyan/lightgrey) for the four-band model.}
\label{SnSi_fatband}
\end{figure}
\begin{figure}[t]
 \includegraphics[width=3.25in]{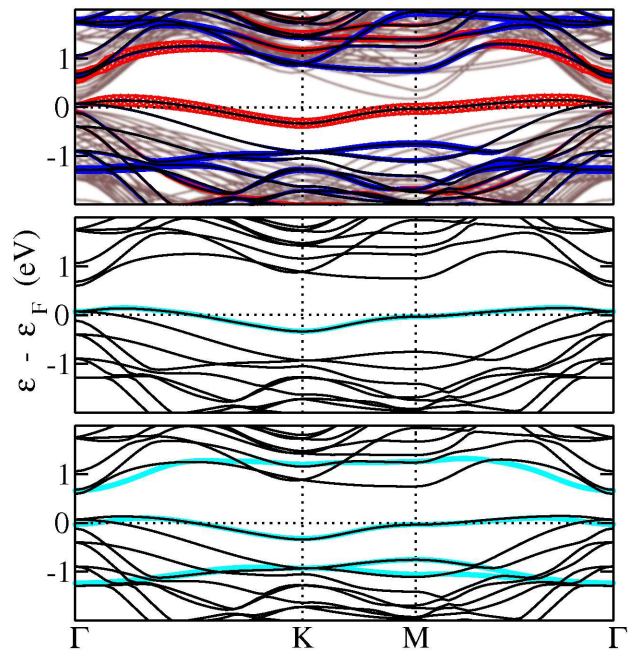}
\caption{(Color online) As Fig.~\ref{SnSi_fatband}, but here for the (flat) 
Sn/Ge(111)-\srt3 system.} 
\label{SnGe_fatband}
\end{figure}
\begin{figure}[t]
   \includegraphics{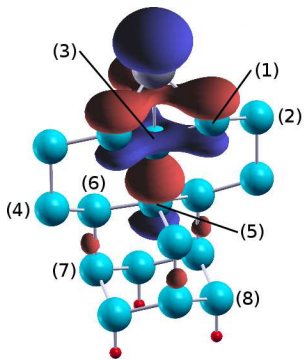}
  \caption{(Color online) Wannier orbital corresponding to the one-band 
  model. The numbers are references for the PBE-GGA DOS plot in Fig.~\ref{surfDOS}}
  \label{wann-single}
  \end{figure}

As seen from the band-structure plots in Fig.~\ref{SnSi_fatband} and 
Fig.~\ref{SnGe_fatband}, the low-energy DOS is associated with a single surface
band within the gap of the bulk system. The so-called fatband resolution for the 
Sn$(5p)$ orbitals, i.e., the identification of the contribution of a given orbital 
character to a Kohn-Sham band at each k point via the width of a broadened coating
'band', underlines the dominating $p_z$ weight on this half-filled surface band. A 
straightforward MLWF construction may be applied to the single surface bands for
both systems. 
The real-space Wannier function (see Fig.~\ref{wann-single}) shows not only the 
$p_z$-like lobe but displays additionally the threefold bonding aspects to the 
substrate atoms. A spread of 14.2 \AA{}$^{2}$ for this MLWF was obtained in the 
case of Sn/Si(111). Albeit a rather similar low-energy state may be identified in 
the flat Sn/Ge(111) system, it has to be noted that conventional DFT has its 
problems in describing germanium systems. For the bulk material, there is a nearly 
vanishing band gap within DFT~\cite{bac85} and additionally relativistic effects on 
the band structure are also important~\cite{glo80,bac85}. Still in our PBE-GGA 
study a single Wannier band may also be extracted in this case 
(see Fig.~\ref{SnGe_fatband}), however contrary to Sn/Si(111) that band touches 
occupied levels at the $\Gamma$ point. 

\begin{figure}[t]
   \includegraphics[width=3.20in]{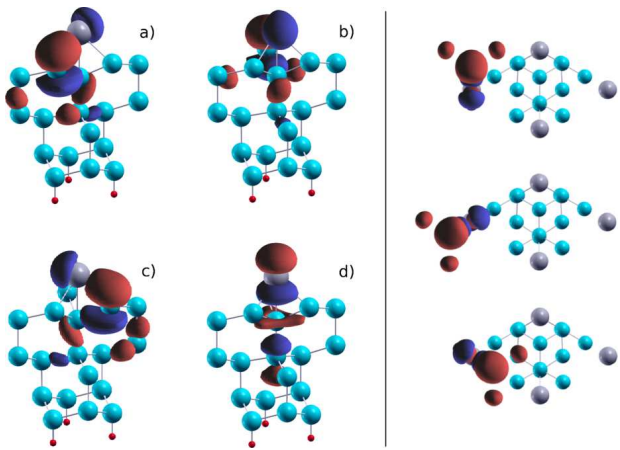}
  \caption{(Color online) Left: Wannier-like orbitals of the four-band model. 
(a-c) $sp^{2}$-like orbitals and (d) p$_{z}$-like orbital.
Right: Top view on the spatial orientations of the three $sp^{2}$-like orbitals 
relative to the \srt3 unit cell.}
  \label{4_band}
  \end{figure}

Due to the substrate geometry and the apparent Sn($5s$) contribution 
at the Fermi level, an extended low-energy modeling based on $sp^2$ hybridized 
orbitals seems even more adequate for these systems. Indeed as shown in 
Figs.~(\ref{SnSi_fatband},~\ref{SnGe_fatband}), the effective bands from the 
corresponding four-band MLWF construction fit well to the full 
Sn($5p$) fatband dispersion (Sn($5s$) has nearly exclusively weight on the band at
the Fermi level).  The chemically more appealing $sp^{2}$+$p_{z}$ viewpoint yields 
three of the four orbitals having a dominant in-plane orientation with an 
120$^{\circ}$ angle between them, while a fourth one is now directly reminiscent of 
the $p_z$ orbital (see  Fig.~\ref{4_band}). This latter Wannier-like orbital has 
indeed major weight on the low-energy band close to $\varepsilon_F$. Compared to 
the one-band case, the spread of the Wannier-like functions are now given as 
13.6 \AA{}$^{2}$ for the $sp^{2}$-like orbitals and 20.8 \AA{}$^{2}$ for the 
remaining $p_z$-like orbital. 

The hopping intergrals for both Wannier constructions are provided
in Tab.~\ref{hoppings_1-4-band}. In the one-band model the nearest-neighbor hopping 
with an absolute value $|t|$$\sim$45 meV is negative in accordance with the 
hole-like dispersion from Fig.~(\ref{SnSi_fatband},~\ref{SnGe_fatband}) around the 
$\Gamma$ point. For the Sn/Ge(111) system the more distant hoppings are slightly 
larger, whereas $|t|$ is somewhat smaller compared to the Sn/Si(111) case. 
\begin{figure}[t]
   \includegraphics[width=2.5in]{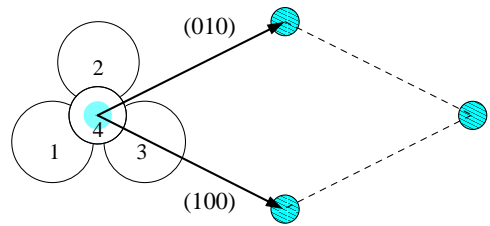}
  \caption{(Color online) Illustration of the Wannier-like orbitals in the
four-band model for the assignement of the hoppings in 
Tab.~\ref{hoppings_1-4-band}. The sketched orbitals 1, 2 and 3 correspond to the 
top, middle and bottom orbitals in the right part of Fig.~\ref{4_band}. 
Orbital number 4 denotes the remaining $p_z$-like level.}
  \label{hopping-pic}
  \end{figure}
The onsite energies in the four-band model yield a crystal-field splitting 
between the $sp^2$-like orbitals and the $p_z$-like one of the order of 1.6 
(0.9) eV for Sn/(Si,Ge)(111). Though some caution has to be taken because of the 
named DFT-Ge problems, this illustrates the more isolated $p_z$-like level for the 
Si substrate and underlines the stronger hybridzation in the case of the 
germanium substrate.

While the minimal one-band model is easily justified, the eligibility 
of the four-band model needs some additional comments. Albeit from the bonding 
character chemically meaningful, the orbital filling for this model amounts to five 
electrons in the present cases, i.e., one more than the half-filled $sp$ valence
of the carbon-group elements (Si,Ge,Sn). Since equipped with similar 
electronegativity, a significant charge transfer in the strongly covalent surface
systems is not expected. Still the five-electron filling may be acceptable as the
Sn locality of the $(sp^2+p_z)$-like orbitals is neither very strong nor demanded
on formal grounds. In addition we find this model more justified than a simpler 
three-band model for $p$-like orbitals only, which would in the end amount also to 
a five-electron filling for three effective bands. Such an extreme filling of 5/6
would render the interpretation in semi-local terms rather difficult, besides the 
inconsistency with the facts of the chemical bonding.
\begingroup
\squeezetable
\begin{table}[t]
\begin{ruledtabular}
\begin{tabular}{l|c|c|c|c}
direction    & 100  & 110   & 200 &  210 \\ \hline
Sn/Si(111)   & 44.6 & -18.4 & 6.7 & -0.8 \\
Sn/Ge(111)   & 43.2 & -23.7 & 7.3 & -1.8 \\
\end{tabular}
\end{ruledtabular}
\vspace*{0.2cm}
\begin{ruledtabular}
\begin{tabular}{c|c c c c|c c c c}
 direction  &     \multicolumn{4}{c|}{ Sn/Si hoppings [meV]}      & \multicolumn{4}{c}{ Sn/Ge hoppings [meV]}\\ \hline
 
              &           -643.8&\phantom{-}452.0&           452.0&           -318.6&          -433.1&\phantom{-}546.4&\phantom{-}546.4&          -358.3\\
000       & \phantom{-}452.0&          -643.8&\phantom{-}452.0&           -318.6&\phantom{-}546.4&          -433.1&\phantom{-}546.4&          -358.3\\
              & \phantom{-}452.0&\phantom{-}452.0&          -634.8&           -318.6&\phantom{-}546.4&\phantom{-}546.4&          -433.1&          -358.3\\  
              &           -318.6&          -318.6&          -318.6& \phantom{-}990.3&          -358.3&          -358.3&          -358.3&\phantom{-}463.2\\ \hline 

              &            -42.6&           -29.9& \phantom{-}29.4&            -26.7&           -61.5&           -37.8& \phantom{-}3.6&        -20.6\\
100       &            -29.9&           -42.6& \phantom{-}29.4&            -26.7&           -37.8&           -61.5& \phantom{-}3.6&        -20.6\\
              &  \phantom{-}74.2& \phantom{-}74.2& \phantom{-}50.8&            -33.1& \phantom{-}33.8& \phantom{-}33.8& \phantom{-}2.9&\phantom{-}70.9\\  
              &  \phantom{-}85.0& \phantom{-}85.0&           -49.1&  \phantom{-}16.7& \phantom{-}54.9& \phantom{-}54.9& \phantom{-}1.6&\phantom{-}29.2\\ \hline 

              &            -0.7&           -5.9&             -8.4&    \phantom{-}1.5&           -15.2&           -19.0&          -22.8&\phantom{-}1.3\\
110       &  \phantom{-}12.1&           -9.5&            -5.9&   \phantom{-}11.3&  \phantom{-}2.9&           -22.8&          -19.0&\phantom{-}14.3\\
              &  \phantom{-}3.8& \phantom{-}12.1&            -0.7&            -24.5&            -10.9& \phantom{-}2.9&           -15.2&           -4.6\\  
              &           -24.5& \phantom{-}11.3&   \phantom{-}1.5&            -9.4&             -4.6& \phantom{-}14.3& \phantom{-}1.3&           -10.3\\ \hline 

              &            -1.2&           -1.4& \phantom{-}4.2&            -1.2&  \phantom{-}3.6&           -4.2& \phantom{-}2.8& \phantom{-}3.6\\
200       &            -1.4&           -1.2& \phantom{-}4.2&            -1.2&           -4.2&  \phantom{-}3.6& \phantom{-}2.8& \phantom{-}3.6\\
              &            -1.6&           -1.6& \phantom{-}8.0&  \phantom{-}1.9&           -2.5&             -2.5& \phantom{-}0.6&\phantom{-}7.3\\  
              &  \phantom{-}5.3& \phantom{-}5.3&           -2.8&            -1.1& \phantom{-}12.6& \phantom{-}12.6& \phantom{-}6.4&        -10.5\\ \hline 

              &   \phantom{-}0.7&  \phantom{-}0.3& \phantom{-}0.9&            -3.3&           -4.3&           -0.8&             -2.5& \phantom{-}0.8\\
210       &   \phantom{-}2.2&  \phantom{-}0.9& \phantom{-}1.3&            -0.5&           -1.9&  \phantom{-}0.2&            -2.7&           -1.1\\
              &             -0.4&            -2.9&            0.0&            -2.6&           -2.4&            -3.7&             -0.9&          -0.7\\  
              &   \phantom{-}0.0&            -3.4& \phantom{-}1.4&  \phantom{-}3.3& \phantom{-}1.8&            -3.0&   \phantom{-}1.5& \phantom{-}2.9\\ 
\end{tabular}
\end{ruledtabular}
\caption{Hopping integrals up to the fourth-nearest neighbours for the 
minimal one-band case (top) and the 4-band ($sp^2$+$p_z$) model (bottom). For the
latter case, the last entries of the given 4$\times$4 matrix are associated with
the $p_z$-like orbital.
Note that the values correspond to the entries of the real-space Wannier(-like) 
hamiltonian, i.e., a minus sign is included.}
\label{hoppings_1-4-band}
\end{table}
\endgroup

\subsection{Three-site Sn unit cell\label{sec:cluster}}

\begin{figure}[b]
 \includegraphics[width=3.25in]{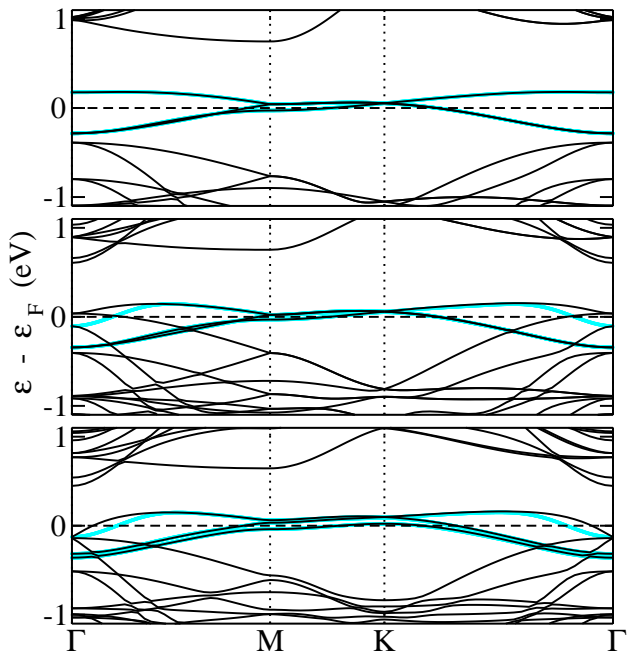}
\caption{(Color online) PBE-GGA band structure (dark) with extracted Wannier(-like) 
bands (cyan/grey) for the 3$\times$3 supercell. Top: planar Sn/Si(111) system. 
Middle: planar Sn/Ge(111) system. Bottom: Sn/Ge(111) with the 2D-1U reconstruction.}
\label{Cluster_bands}
\end{figure}

It is believed from experimental studies that the structural ground state of the 
Sn/Ge(111) system corresponds to a two-down-one-up (2D-1U) distortion of the Sn 
submonolayer~\cite{pul06}. To be able to account for this reconstruction 
(cf. Fig.~\ref{structure}), we thus performed electronic-structure calculations 
using a 3$\times$3 supercell with a difference $\Delta$=0.32 \AA{} between the 
up-down Sn positions~\cite{gir00}, which we applied to our PBE-GGA structurally
relaxed unit cell with the planar submonolayer. In these extended supercell 
calculations only one orbital per Sn adatom is included in the subsequent MLWF 
construction. This results in a Kohn-Sham-Wannier hamiltonian which corresponds to 
a single-orbital problem on a three-site triangular cluster. 
For comparison, we derived such enlarged hamiltonians also for the flat systems via
corresponding larger unit-cell calculations for planar Sn/(Si,Ge)(111). The 
resulting Wannier-like bands are shown in Fig.~\ref{Cluster_bands}. Whereas the 
bands again fit exactly in the Sn/Si(111) case, for the distorted Sn/Ge(111) 
surface a shift towards lower lying bands at the $\Gamma$-point is visible. The
latter feature is again due to the band hybridizations already observed in the
single-site Sn unit cell. It may be observed that the 2D-1U surface reconstruction
leads to a small splitting of the low-energy bands, affecting mainly the occupied 
part of these states. 

Besides the study of the influence of apparent surface reconstructions, the 
investigation of the magnetic behavior seems a most important endeavor. 
As already outlined in the introduction, the quasi-2D triangular Sn submonolayer
might be a realistic case for the application of model ideas discussed in the 
context of 2D quantum magnetism. Therefore we investigated possible magnetic 
orderings within PBE-GGA, especially for the Sn/Si(111) system where intricate 
ordering patterns are heavily discussed~\cite{mod07,pro07}.
Although (finite-temperature) magnetism would be very interesting, one has
to keep in mind that the constituents (Sn,Si,Ge) are no high-susceptible
magnetic materials. The $sp$-bonding (involving large principal quantum numbers) 
with filled $d$ states renders magnetism from a chemical point of view questionable.
Furthermore the rather large Sn-Sn nearest-neighbor distance of 
$\sim$6.7\AA$\,$ asks for a robust exchange path to facilitate the appearance of
long-range order. Still the low-dimensional 2D-character might be sufficient to
induce delicate magnetic behavior. Note also that the high DOS close to the Fermi
level (cf. Fig.~\ref{surfDOS}) might give rise to flat-band 
ferromagnetism~\cite{mie91,tas92}.

\begin{table}[b]
\begin{ruledtabular}
\begin{tabular}{l|r|r|r}
magnetic ordering     & E [meV] & M [$\mu_{\rm B}$] & M$_{\rm Sn}$ 
[$\mu_{\rm B}$]\\ \hline 
ferromagnetic                     &  1.5  & 0.80 & 0.031 \\
collinear ferrimagnetic           & -1.6  & 0.44 & 0.055 \\
120$^{\circ}$ antiferromagnetic   & -3.6  & 0.00 & 0.058 \\
\end{tabular}
\end{ruledtabular}
\caption{
Comparison of the different magnetic orderings in the flat Sn submonolayer of
the \srt3 surface with the Si substrate. The energies $E$ and total moments $M$ 
are given with respect to the nonmagnetic structure in PBE-GGA and correspond to
an enlarged \rt3 unit cell. This cell incorporates 66 atoms, namely 54 Si, 3 Sn 
and 9 H atoms (for saturation of the bottom bulk-like Si layer).}
\label{magtab}
\end{table}
The results of our investigation of different magnetic orderings on planar
Sn/Si(111)~\rt3 are summarized in Tab.~\ref{magtab}. The calculations show 
that especially the ferromagnetic (FM) order is rather intriguing. One may 
stabilize a FM solution for Sn/Si(111), resembling previous 
work by Profetta and Tosatti~\cite{pro07}, however this state is energetically 
unfavorable compared to the nonmagnetic (NM) solution. The corresponding local energy
minimum of this FM state appears to be rather flat, thus already small disturbances
drive the DFT self-consistency cycle towards the NM state. This was confirmed
within all three utilized band-structure codes, i.e., MBPP, CP-PAW and VASP, whereby
the Sn-substrate distance does not influence this qualitative result. Note the
rather small local $M_{\rm Sn}$ moment compared to total FM moment of the supercell.
Thus this metastable FM state is far from being originated from pure local Sn
moments, but has significant nonlocal character. Since 
collinear AFM order is impossible due to frustration on the undistorted
lattice, a collinear ferrimagnetic ordering (two up spins and one down spin on the 
minimal triangle) as well as the in-plane noncollinear 120$^{\circ}$-AFM state were 
investigated. Indeed, both latter ordering patterns are found to be stable with 
respect to the nonmagnetic solution, with the lowest total energy for the 
120$^{\circ}$-AFM ordering. Though the local Sn magnetic moments $M_{\rm Sn}$
are only of the order of $\sim$0.06 $\mu_{\rm B}$ within PBE-GGA, detailed
convergence studies elucidated nonetheless their nonzero value. Note that these 
local $M_{\rm Sn}$ are supplemented by additonal spin-polarisation on the remaining 
sites and the interstitial contribute. Hence again the picture of strictly 
localized magnetism is not appropriate on the weakly-correlated modeling level, but
still local moments with small AFM exchange may exist in these Sn submonolayers. 
Those may possibly give rise to spin-liquid physics or eventual magnetic 
long-range order also in the correlated regime~\cite{san99}. 
 
Similar magnetic PBE-GGA studies for a model {\sl planar} Sn/Ge(111)~\rt3 surface 
did not result in magnetic long-range order. The stronger hybridization
of the Ge($4s4p$) states with tin should generally weaken the magnetic tendencies
compared to the silicon substrate. We did not investigate the more realistic 2D-1U 
resonstructed structure in this matter, which is left for further studies.

\section{Investigation of electronic correlations}

In order to take the principal effect of electronic correlations into account, we
concentrate in the following on the realistic one-band models derived in 
section~\ref{sec:single} and their cluster extension from section~\ref{sec:cluster}.
Furthermore we restrict the investigations to paramagnetic modelings, i.e. do not
cover possible magnetic orderings. Since it became clear from the last
section~\ref{sec:cluster} that the energy scale for magnetic long-range order
is rather small, such an analysis shall be postponed to future studies.

\subsection{LDA+DMFT(QMC) study}

\begin{figure}[t]
\includegraphics*[clip,width=3.00in]{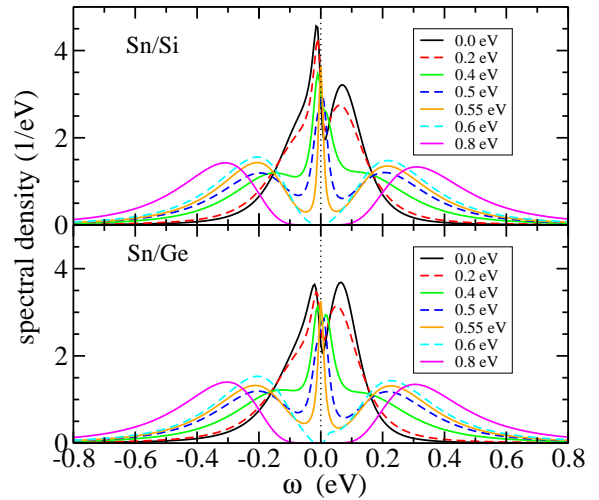}
\caption{(Color online) Realistic single-site DMFT(QMC) \newline results for the planar 
one-band models at $\beta$=200 eV$^{-1}$ \newline ($T$$\sim$ 58K).}
\label{dmft_1site}
\end{figure}

By combining our Wannier hamiltonians with the DMFT framework we are in the 
position to reveal the spectral function of the surface systems in the interacting
regime. Due to the small bandwidth of $W$$\sim$0.4 eV, already small absolute 
values for the Hubbard $U$ may introduce strong correlation effects. Although the 
low-energy bands of the discussed systems are mainly composed of Sn($5s$,$5p$) 
states, an onsite Coulomb interaction of this order of magnitude may very well be 
reasonable.
Note that also the screening capabilities because of the semiconducting substrate 
are limited. The results of the paramagnetic single-site DMFT(QMC) calculations 
for the realistic one-band model are displayed in Fig.~\ref{dmft_1site} for 
different choices of the Hubbard $U$. Of course, due to the simplicity of the 
modeling the spectral function (finally obtained via the maximum-entropy method) 
follows the usual behavior with increasing $U$, i.e., a low-energy band narrowing 
with the additional appearance of Hubbard-band features at higher energies takes 
place. The Mott transition is reached at a critical value of about $U_c$$\sim$0.6 eV 
for the Sn/Si(111) surface systems (with a somewhat larger value for planar 
Sn/Ge(111)). Since this transition shows a weak first-order character in the
calculations (see also section~\ref{sec-risb}) critical interactions 
$U_{c1}$, $U_{c2}$ govern this regime, respectively. However in the present study, 
due to the simpicity of the one-band modeling, we did not map out the respective 
hysteresis loops.
\begin{figure}[t]
\includegraphics*[clip,width=3.00in]{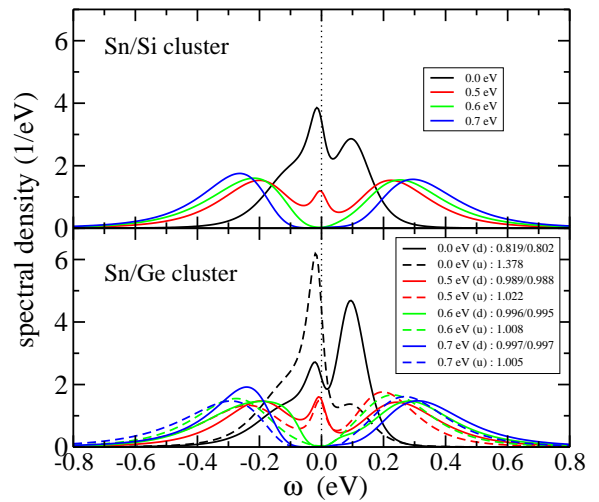}
\caption{(Color online) Realistic cellular-cluster DMFT(QMC) results for planar
Sn/Si(111) and 2D-1U Sn/Ge(111). The up/down tags refer to the inequivalent atomic 
positions and the numbers yield the orbital occupation. Note that albeit slightly
different by symmetry, we only show the averaged spectral function for the two 'down'
atoms (occupations are site selective though). Calculations were performed at 
$\beta$=70 eV$^{-1}$ ($T$$\sim$ 166K).}
\label{dmft_cluster}
\end{figure}
The energy gap within the insulating Mott state is of the order of $\sim$0.2 eV 
for $U$$\sim$0.7 eV with Hubbard bands at around 0.3 eV above and below $E_F$ for
both Sn/(Si,Ge)(111) systems. The position of our Hubbard excitations is roughly
in line with the region of increased spectral-weight transfer measured in 
photoemission experiments~\cite{uhr00,mod07} below 30K. Hence a rather small
Hubbard $U$, slightly larger than the single bandwidth, is sufficient to drive the
Sn-dominated surface band Mott insulating. This value is however not very suprising
since from the involved quantum numbers of the respective states no large value
of $U$ (i.e. as for transition-metal and/or $f$ systems) is expected.

In order to study the importance of intersite self-energy effects, especially for
the distorted Sn/Ge(111) case, we also performed cluster DMFT 
(CDMFT) (for recent reviews see e.g. Ref.~[\onlinecite{lic02,bir04,mai05}]) 
computations within the cellular cluster framework for the basic Sn triangle in the
submonolayer. The resulting spectral functions are shown in 
Fig.~\ref{dmft_cluster}. Note that we employed a somewhat higher temperature within 
the QMC solver because of the larger numerical effort in the cluster framework. For 
the planar Sn/Si(111) case the main differences to the single-site results are 
given by a slightly smaller $U_c$$\sim$0.55 eV, smaller energy gap and Hubbard 
bands in some closer range to the low-energy region. We checked that those 
observations are not only due to the different temperatures of our computations. 
One may also observe
a stronger asymmetry in the spectrum up and below the Fermi level, with some
stronger reduction of low-energy spectral weight below $E_F$. For the site-resolved
spectral function of the Sn/Ge(111) 3$\times$3 system in the distorted 2D-1U 
structure it is first important to remark that the two downwards shifted Sn atoms 
are inequivalent by symmetry due to different hybridization with the 'up' atom even
assuming the same height for the 'down' 2D atoms~\cite{tej08,gor09} (see 
section~\ref{sec-risb}). The latter are 
moreover less occupied than the upwards shifted Sn atom~\cite{avi99}. With 
increasing $U$, this filling inbalance because of the distortion-induced 
crystal-field shifts becomes smaller and is compensated at the Mott transition
(see Fig.~\ref{dmft_cluster}). However a slightly larger critical $U_c$ compared
to the planar cluster for Sn/Ge(111) is necessary to reach this transition (more
or less equivalent to the critical $U$ within the planar single-site DMFT). The
above noted spectral asymmetry in the occupied and unoccupied part is even larger in the
distorted case.

\subsection{LDA+RISB study\label{sec-risb}}

\begin{figure}[t]
\includegraphics*[width=3.00in]{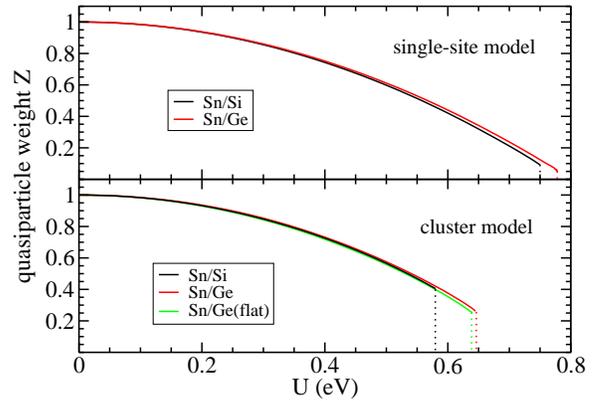}
\caption{(Color online) RISB results for the quasiparticle weight $Z$ for the 
planar single-site (top) and the cellular-cluster (bottom) models.}
\label{RISB}
\end{figure}

\begin{figure}[t]
\includegraphics*[width=3.00in]{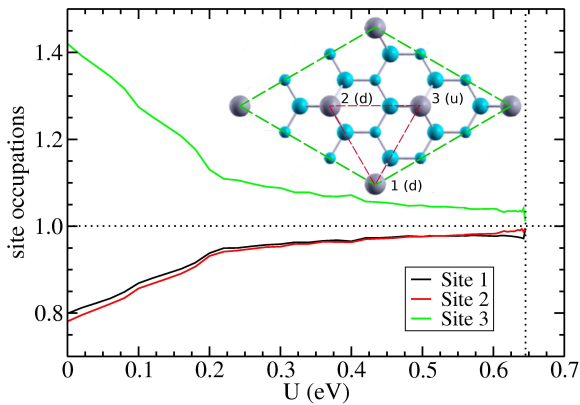}
\caption{(Color online) Orbital occupations of the different Sn atoms in the 2D-1U
structure for Sn/Ge(111) obtained from the cellular cluster calculation with
varying $U$. Inset: Designation of the different Sn atoms in the cellular 
cluster description of Sn/Ge(111) from a top view on the surface. The (red/grey)
dashed lines mark the cluster utilized for the cellular modeling.}
  \label{siteocc}
\end{figure}

In addition to DMFT(QMC) computations we have performed slave-boson 
calculations within the RISB framework to further verify our results and to achieve 
a better resolution of the qualitative differences in values for the critical $U$ 
in the various cases. Furthermore we also want to elucidate the intersite spin 
correlations in the cellular cluster modeling. 
The RISB results for the quasiparticle weight $Z$ are shown in Fig.~\ref{RISB}. 
In the one-band case, the RISB method yields for both flat systems a first-order
transition from the paramagnetic metal to the paramagnetic insulator. The 
corresponding critical $U$ values are given by $U_c$$\sim$0.75 eV (0.78 eV) for
for Sn/(Si,Ge)(111) \srt3, hence show the same qualitative trend as DMFT(QMC). A
somewhat larger absolute value within RISB is understandable from the saddle-point
approximation which is identical to the neglection of quantum fluctuations. The
cellular-cluster investigations again verify the reduced critical Hubbard $U$ values
(similar to the ones from DMFT(QMC)) due to an increase in the correlation 
strength via the inclusion of the nearest-neighbor self energies. Also the 
enhanced $U_c$ for the distorted 2D-1U structure of Sn/Ge(111) is a solid result. 
Note that the first-order character of the transition is strenghtened in the cluster
modeling.
To allow for a commensurable Mott transition, the correlations have to drive 
additional charge transfers between the Sn site in the 2D-1U structure. The 
therefore enhanced charge fluctuations in the latter case may thus account for the 
larger $U_c$~\cite{flo01}. In Fig.~\ref{siteocc} we show the continuous development 
of the individual site fillings with increasing $U$ for the 2D-1U structure. The
corresponding labeling of the different Sn atoms is given in the inset of
Fig.~\ref{siteocc}.
As it is clearly seen, the originally enhanced occupation of the low-energy orbital
for the upwards shifted Sn atom on the cluster becomes reduced with increasing $U$, 
while the downwards shifted ones gain electron filling in their respective orbitals.
This is in line with the results obtained from the more elaborate QMC solver to DMFT
(cf. Fig.~\ref{dmft_cluster}).

\begin{figure}[t]
\includegraphics*[width=3.00in]{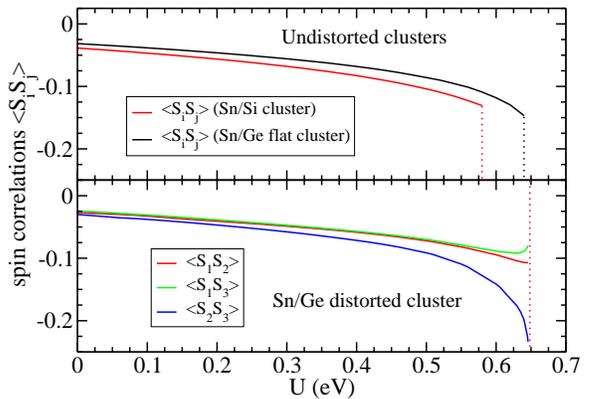}
\caption{(Color online) Spin correlation functions for the Sn/Si and the 
Sn/Ge distorted cluster obtained from the RISB study. Top: planar systems, 
bottom: distorted 2D-1U Sn/Ge(111) system. The vertical line marks the 
critical $U$ value, respectively.}
\label{spincorr}
\end{figure}

In order to provide some insight into the magnetic behavior with taking into account
electronic correlations, Fig.~\ref{spincorr} displays the local spin correlations 
$\langle {\bf S}_i{\bf S}_j\rangle$ between the Sn atoms from the low-energy 
modeling. It is seen that the spin correlations are always negative, i.e., of AFM 
character, as expected by considering the superexchange induced via $U$. Naturally,
the degree of magnetic behavior is therewith increased compared to the noninteracting
case. The comparison of the two different substrates with the planar geometry of 
the Sn submonolayer exhibits the stronger magnetic correlations within the 
Sn/Si(111) system. Thus the qualitative result obtained from the weakly-correlated 
PBE-GGA modeling extends to the strongly correlated treatment. In the 2D-1U 
structure of Sn/Ge(111), the values of $\langle {\bf S}_i{\bf S}_j\rangle$ for the 
now different Sn-Sn pairs show interesting behavior. Somewhat counterintuitive, the 
spin correlations between the two up-down pairs, here denoted Sn(1)-Sn(3) and 
Sn(2)-Sn(3), scale rather differently with increasing $U$. While for the 
Sn(2)-Sn(3) pair  $\langle {\bf S}_i{\bf S}_j\rangle$ is most negative, for the 
Sn(1)-Sn(3) pair the spin correlations become nonmonotonic close to $U_c$, 
anticipating tendencies to eventual FM coupling. The remaining Sn(1)-Sn(2) pair of 
the two down atoms scales inbetween these functions. This result can be explained 
by the fact that the absolute value of the hopping amplitude $t$ is maximum between 
the Sn(2)-Sn(3) pair and is accordingly weakened between Sn(1)-Sn(3). Because of 
the upward shift of the Sn(3) atom, the lobe from its effective Wannier orbital 
pointing towards Sn(2) achieves a wider range, leading to stronger overlap with the 
Wannier orbital of Sn(2). 

\section{Summary}

We have presented a detailed modeling of the Sn/(Si,Ge)(111) surface systems using
state-of-the-art band-structure methods combined with many-body techniques. In 
addition to previous DFT investigations we extracted realistic one- and four-band
Wannier(-like) hamiltonians that are capable of describing the electronic structure
close to the Fermi level. The physics of the minimal one-band model is of course 
restricted, however due to the prominent Sn-dominated half-filled surface band at 
$\varepsilon_F$, especially for Sn/Si(111) \srt3 this is believed to be an adequate
modeling viewpoint. The atomistic description of the Sn/Ge(111) system still raises 
some questions concerning the apparent limitations of simplified 
exchange-correlation functionals in this case. Nonetheless on the present level of 
investigation and comparison, the PBE-GGA perspective appears sufficient to reveal 
the essential differences between both surface systems. The hybridization of the Sn 
submonolayer with the substrate is stronger in the case of Ge, leading there to a 
more intriguing entanglement of that layer with the supporting atoms below.
Consequently, structural reconstructions are more likely and indeed take place as
verified in experimental studies.

The magnetic behavior in these systems renders them most fascinating from a
fundamental physics point of view in terms of possible low-dimensional quantum 
magnetism. The (non)collinear spin-polarized DFT calculations revealed that there 
is indeed the chance for intricate magnetic orderings. In this respect we found the 
120$^\circ$-AFM structure to be the magnetic ground state in the weakly-correlated 
description. Further theoretical studies, including the effect of true many-body
correlations on the magnetism are important and should be motivated by the present 
work. Although the resulting moments are rather small, maybe advanced experimental 
techniques are capable of exploring these small energy scales.

Using DMFT(QMC) and RISB techniques, a moderate value of the order of 
$U$$\sim$0.5-0.6 eV was found to govern the systems from a one-band Hubbard-like 
model in a strongly-correlated limit. Because of the small low-energy bandwidth 
it is reasonable 
to expect electronic correlations to be important, however sole local Coulomb 
interactions are surely limited in the description of these systems based on 
carbon-group elements. Yet already on this rather simple level a delicate interplay 
between the structural and electronic degrees of freedom, anticipated in other
model(-like) frameworks~\cite{san99,flo01}, was verified on the grounds of an
LDA+DMFT approach. 
Future studies now have to concentrate on including the substrate degrees of 
freedom explicitly in the modeling of the interacting electronic structure, perhaps
even allowing for the dynamical structural reconstructions needed for Sn/Ge(111).

\acknowledgements

We thank C.~Piefke, L.~Boehnke, J.~Wiebe and S.~Modesti for helpful discussions. Financial 
support was provided by the SFB668 and the NANOSPINTRONICS Cluster of Excellence.
Computations were performed at the Linux-Cluster of the RRZ at the University of
Hamburg as well as the North-German Supercomputing Alliance (HRLN).

\bibliographystyle{apsrev}
\bibliography{bibextra}

\begin{thebibliography}{53}
\expandafter\ifx\csname natexlab\endcsname\relax\def\natexlab#1{#1}\fi
\expandafter\ifx\csname bibnamefont\endcsname\relax
  \def\bibnamefont#1{#1}\fi
\expandafter\ifx\csname bibfnamefont\endcsname\relax
  \def\bibfnamefont#1{#1}\fi
\expandafter\ifx\csname citenamefont\endcsname\relax
  \def\citenamefont#1{#1}\fi
\expandafter\ifx\csname url\endcsname\relax
  \def\url#1{\texttt{#1}}\fi
\expandafter\ifx\csname urlprefix\endcsname\relax\def\urlprefix{URL }\fi
\providecommand{\bibinfo}[2]{#2}
\providecommand{\eprint}[2][]{\url{#2}}

\bibitem[{\citenamefont{Carpinelli et~al.}(1996)\citenamefont{Carpinelli,
  Weitering, Plummer, and Stumpf}}]{car96}
\bibinfo{author}{\bibfnamefont{J.~M.} \bibnamefont{Carpinelli}},
  \bibinfo{author}{\bibfnamefont{H.~H.} \bibnamefont{Weitering}},
  \bibinfo{author}{\bibfnamefont{E.~W.} \bibnamefont{Plummer}},
  \bibnamefont{and} \bibinfo{author}{\bibfnamefont{R.}~\bibnamefont{Stumpf}},
  \bibinfo{journal}{Nature} \textbf{\bibinfo{volume}{381}},
  \bibinfo{pages}{398} (\bibinfo{year}{1996}).

\bibitem[{\citenamefont{Weitering et~al.}(1997)\citenamefont{Weitering, Shi,
  Johnson, Chen, DiNardo, and Kempa}}]{wei97}
\bibinfo{author}{\bibfnamefont{H.~H.} \bibnamefont{Weitering}},
  \bibinfo{author}{\bibfnamefont{X.}~\bibnamefont{Shi}},
  \bibinfo{author}{\bibfnamefont{P.~D.} \bibnamefont{Johnson}},
  \bibinfo{author}{\bibfnamefont{J.}~\bibnamefont{Chen}},
  \bibinfo{author}{\bibfnamefont{N.~J.} \bibnamefont{DiNardo}},
  \bibnamefont{and} \bibinfo{author}{\bibfnamefont{K.}~\bibnamefont{Kempa}},
  \bibinfo{journal}{Phys. Rev. Lett.} \textbf{\bibinfo{volume}{78}},
  \bibinfo{pages}{1331} (\bibinfo{year}{1997}).

\bibitem[{\citenamefont{Zhang et~al.}(2010)\citenamefont{Zhang, Cheng, Li, Sun,
  Wang, Zhu, He, Wang, Ma, Chen et~al.}}]{zha10}
\bibinfo{author}{\bibfnamefont{T.}~\bibnamefont{Zhang}},
  \bibinfo{author}{\bibfnamefont{P.}~\bibnamefont{Cheng}},
  \bibinfo{author}{\bibfnamefont{W.-J.} \bibnamefont{Li}},
  \bibinfo{author}{\bibfnamefont{Y.-J.} \bibnamefont{Sun}},
  \bibinfo{author}{\bibfnamefont{G.}~\bibnamefont{Wang}},
  \bibinfo{author}{\bibfnamefont{X.-G.} \bibnamefont{Zhu}},
  \bibinfo{author}{\bibfnamefont{K.}~\bibnamefont{He}},
  \bibinfo{author}{\bibfnamefont{L.}~\bibnamefont{Wang}},
  \bibinfo{author}{\bibfnamefont{X.}~\bibnamefont{Ma}},
  \bibinfo{author}{\bibfnamefont{X.}~\bibnamefont{Chen}}, \bibnamefont{et~al.},
  \bibinfo{journal}{Nature Physics} \textbf{\bibinfo{volume}{6}},
  \bibinfo{pages}{104} (\bibinfo{year}{2010}).

\bibitem[{\citenamefont{Modesti et~al.}(2007)\citenamefont{Modesti, Petaccia,
  Ceballos, Vobornik, Panaccione, Rossi, Ottaviano, Larciprete, Lizzit, and
  Goldoni}}]{mod07}
\bibinfo{author}{\bibfnamefont{S.}~\bibnamefont{Modesti}},
  \bibinfo{author}{\bibfnamefont{L.}~\bibnamefont{Petaccia}},
  \bibinfo{author}{\bibfnamefont{G.}~\bibnamefont{Ceballos}},
  \bibinfo{author}{\bibfnamefont{I.}~\bibnamefont{Vobornik}},
  \bibinfo{author}{\bibfnamefont{G.}~\bibnamefont{Panaccione}},
  \bibinfo{author}{\bibfnamefont{G.}~\bibnamefont{Rossi}},
  \bibinfo{author}{\bibfnamefont{L.}~\bibnamefont{Ottaviano}},
  \bibinfo{author}{\bibfnamefont{R.}~\bibnamefont{Larciprete}},
  \bibinfo{author}{\bibfnamefont{S.}~\bibnamefont{Lizzit}}, \bibnamefont{and}
  \bibinfo{author}{\bibfnamefont{A.}~\bibnamefont{Goldoni}},
  \bibinfo{journal}{Phys. Rev. Lett.} \textbf{\bibinfo{volume}{98}},
  \bibinfo{pages}{126401} (\bibinfo{year}{2007}).

\bibitem[{\citenamefont{Morikawa et~al.}(2002)\citenamefont{Morikawa, Matsuda,
  and Hasegawa}}]{mor02}
\bibinfo{author}{\bibfnamefont{H.}~\bibnamefont{Morikawa}},
  \bibinfo{author}{\bibfnamefont{I.}~\bibnamefont{Matsuda}}, \bibnamefont{and}
  \bibinfo{author}{\bibfnamefont{S.}~\bibnamefont{Hasegawa}},
  \bibinfo{journal}{Phys. Rev. B} \textbf{\bibinfo{volume}{65}},
  \bibinfo{pages}{201308} (\bibinfo{year}{2002}).

\bibitem[{\citenamefont{Uhrberg et~al.}(2000)\citenamefont{Uhrberg, Zhang,
  Balasubramanian, Jemander, Lin, and Hansson}}]{uhr00}
\bibinfo{author}{\bibfnamefont{R.~I.~G.} \bibnamefont{Uhrberg}},
  \bibinfo{author}{\bibfnamefont{H.~M.} \bibnamefont{Zhang}},
  \bibinfo{author}{\bibfnamefont{T.}~\bibnamefont{Balasubramanian}},
  \bibinfo{author}{\bibfnamefont{S.~T.} \bibnamefont{Jemander}},
  \bibinfo{author}{\bibfnamefont{N.}~\bibnamefont{Lin}}, \bibnamefont{and}
  \bibinfo{author}{\bibfnamefont{G.~V.} \bibnamefont{Hansson}},
  \bibinfo{journal}{Phys. Rev. B} \textbf{\bibinfo{volume}{62}},
  \bibinfo{pages}{8082} (\bibinfo{year}{2000}).

\bibitem[{\citenamefont{Charrier et~al.}(2001)\citenamefont{Charrier,
  P\'{e}rez, Thibaudau, Debever, Ortega, Flores, and Themlin}}]{cha01}
\bibinfo{author}{\bibfnamefont{A.}~\bibnamefont{Charrier}},
  \bibinfo{author}{\bibfnamefont{R.}~\bibnamefont{P\'{e}rez}},
  \bibinfo{author}{\bibfnamefont{F.}~\bibnamefont{Thibaudau}},
  \bibinfo{author}{\bibfnamefont{J.-M.} \bibnamefont{Debever}},
  \bibinfo{author}{\bibfnamefont{J.}~\bibnamefont{Ortega}},
  \bibinfo{author}{\bibfnamefont{F.}~\bibnamefont{Flores}}, \bibnamefont{and}
  \bibinfo{author}{\bibfnamefont{J.-M.} \bibnamefont{Themlin}},
  \bibinfo{journal}{Phys. Rev. B} \textbf{\bibinfo{volume}{64}},
  \bibinfo{pages}{115407} (\bibinfo{year}{2001}).

\bibitem[{\citenamefont{Avila et~al.}(1999)\citenamefont{Avila, Mascaraque,
  Michel, Asensio, LeLay, Ortega, P\'{e}rez, and Flores}}]{avi99}
\bibinfo{author}{\bibfnamefont{J.}~\bibnamefont{Avila}},
  \bibinfo{author}{\bibfnamefont{A.}~\bibnamefont{Mascaraque}},
  \bibinfo{author}{\bibfnamefont{E.~G.} \bibnamefont{Michel}},
  \bibinfo{author}{\bibfnamefont{M.~C.} \bibnamefont{Asensio}},
  \bibinfo{author}{\bibfnamefont{G.}~\bibnamefont{LeLay}},
  \bibinfo{author}{\bibfnamefont{J.}~\bibnamefont{Ortega}},
  \bibinfo{author}{\bibfnamefont{R.}~\bibnamefont{P\'{e}rez}},
  \bibnamefont{and} \bibinfo{author}{\bibfnamefont{F.}~\bibnamefont{Flores}},
  \bibinfo{journal}{Phys. Rev. Lett.} \textbf{\bibinfo{volume}{82}},
  \bibinfo{pages}{442} (\bibinfo{year}{1999}).

\bibitem[{\citenamefont{Pulci et~al.}(2006)\citenamefont{Pulci, Marsili, Gori,
  Palummo, Cricenti, Bechstedt, and del Sole}}]{pul06}
\bibinfo{author}{\bibfnamefont{O.}~\bibnamefont{Pulci}},
  \bibinfo{author}{\bibfnamefont{M.}~\bibnamefont{Marsili}},
  \bibinfo{author}{\bibfnamefont{P.}~\bibnamefont{Gori}},
  \bibinfo{author}{\bibfnamefont{M.}~\bibnamefont{Palummo}},
  \bibinfo{author}{\bibfnamefont{A.}~\bibnamefont{Cricenti}},
  \bibinfo{author}{\bibfnamefont{F.}~\bibnamefont{Bechstedt}},
  \bibnamefont{and} \bibinfo{author}{\bibfnamefont{R.}~\bibnamefont{del Sole}},
  \bibinfo{journal}{Appl. Phys. A} \textbf{\bibinfo{volume}{85}},
  \bibinfo{pages}{361} (\bibinfo{year}{2006}).

\bibitem[{\citenamefont{Cort\'{e}s et~al.}(2006)\citenamefont{Cort\'{e}s,
  Tejeda, Lobo, Didiot, Kierren, Malterre, Michel, and Mascaraque}}]{cor06}
\bibinfo{author}{\bibfnamefont{R.}~\bibnamefont{Cort\'{e}s}},
  \bibinfo{author}{\bibfnamefont{A.}~\bibnamefont{Tejeda}},
  \bibinfo{author}{\bibfnamefont{J.}~\bibnamefont{Lobo}},
  \bibinfo{author}{\bibfnamefont{C.}~\bibnamefont{Didiot}},
  \bibinfo{author}{\bibfnamefont{B.}~\bibnamefont{Kierren}},
  \bibinfo{author}{\bibfnamefont{D.}~\bibnamefont{Malterre}},
  \bibinfo{author}{\bibfnamefont{E.~G.} \bibnamefont{Michel}},
  \bibnamefont{and}
  \bibinfo{author}{\bibfnamefont{A.}~\bibnamefont{Mascaraque}},
  \bibinfo{journal}{Phys. Rev. Lett.} \textbf{\bibinfo{volume}{96}},
  \bibinfo{pages}{126103} (\bibinfo{year}{2006}).

\bibitem[{\citenamefont{Colonna et~al.}(2008)\citenamefont{Colonna, Ronci,
  Cricenti, and LeLay}}]{col08}
\bibinfo{author}{\bibfnamefont{S.}~\bibnamefont{Colonna}},
  \bibinfo{author}{\bibfnamefont{F.}~\bibnamefont{Ronci}},
  \bibinfo{author}{\bibfnamefont{A.}~\bibnamefont{Cricenti}}, \bibnamefont{and}
  \bibinfo{author}{\bibfnamefont{G.}~\bibnamefont{LeLay}},
  \bibinfo{journal}{Phys. Rev. Lett.} \textbf{\bibinfo{volume}{101}},
  \bibinfo{pages}{186102} (\bibinfo{year}{2008}).

\bibitem[{\citenamefont{Morikawa et~al.}(2008)\citenamefont{Morikawa, Jeong,
  and Yeom}}]{mor08}
\bibinfo{author}{\bibfnamefont{H.}~\bibnamefont{Morikawa}},
  \bibinfo{author}{\bibfnamefont{S.}~\bibnamefont{Jeong}}, \bibnamefont{and}
  \bibinfo{author}{\bibfnamefont{H.~W.} \bibnamefont{Yeom}},
  \bibinfo{journal}{Phys. Rev. B} \textbf{\bibinfo{volume}{78}},
  \bibinfo{pages}{245307} (\bibinfo{year}{2008}).

\bibitem[{\citenamefont{Santoro et~al.}(1999)\citenamefont{Santoro, Scandolo,
  and Tosatti}}]{san99}
\bibinfo{author}{\bibfnamefont{G.}~\bibnamefont{Santoro}},
  \bibinfo{author}{\bibfnamefont{S.}~\bibnamefont{Scandolo}}, \bibnamefont{and}
  \bibinfo{author}{\bibfnamefont{E.}~\bibnamefont{Tosatti}},
  \bibinfo{journal}{Phys. Rev. B} \textbf{\bibinfo{volume}{59}},
  \bibinfo{pages}{1891} (\bibinfo{year}{1999}).

\bibitem[{\citenamefont{Profeta et~al.}(2000)\citenamefont{Profeta, Continenza,
  Ottaviano, Mannstadt, and Freeman}}]{pro00}
\bibinfo{author}{\bibfnamefont{G.}~\bibnamefont{Profeta}},
  \bibinfo{author}{\bibfnamefont{A.}~\bibnamefont{Continenza}},
  \bibinfo{author}{\bibfnamefont{L.}~\bibnamefont{Ottaviano}},
  \bibinfo{author}{\bibfnamefont{W.}~\bibnamefont{Mannstadt}},
  \bibnamefont{and} \bibinfo{author}{\bibfnamefont{A.~J.}
  \bibnamefont{Freeman}}, \bibinfo{journal}{Phys. Rev. B}
  \textbf{\bibinfo{volume}{62}}, \bibinfo{pages}{1556} (\bibinfo{year}{2000}).

\bibitem[{\citenamefont{Flores et~al.}(2001)\citenamefont{Flores, Ortega,
  P\'{e}rez, Charrier, Thibaudau, Debever, and Themlin}}]{flo01}
\bibinfo{author}{\bibfnamefont{F.}~\bibnamefont{Flores}},
  \bibinfo{author}{\bibfnamefont{J.}~\bibnamefont{Ortega}},
  \bibinfo{author}{\bibfnamefont{R.}~\bibnamefont{P\'{e}rez}},
  \bibinfo{author}{\bibfnamefont{A.}~\bibnamefont{Charrier}},
  \bibinfo{author}{\bibfnamefont{F.}~\bibnamefont{Thibaudau}},
  \bibinfo{author}{\bibfnamefont{J.-M.} \bibnamefont{Debever}},
  \bibnamefont{and} \bibinfo{author}{\bibfnamefont{J.-M.}
  \bibnamefont{Themlin}}, \bibinfo{journal}{Prog. Surf. Sci}
  \textbf{\bibinfo{volume}{67}}, \bibinfo{pages}{299} (\bibinfo{year}{2001}).

\bibitem[{\citenamefont{{de~Gironcoli}
  et~al.}(2000)\citenamefont{{de~Gironcoli}, Scandolo, Ballabio, Santoro, and
  Tosatti}}]{gir00}
\bibinfo{author}{\bibfnamefont{S.}~\bibnamefont{{de~Gironcoli}}},
  \bibinfo{author}{\bibfnamefont{S.}~\bibnamefont{Scandolo}},
  \bibinfo{author}{\bibfnamefont{G.}~\bibnamefont{Ballabio}},
  \bibinfo{author}{\bibfnamefont{G.}~\bibnamefont{Santoro}}, \bibnamefont{and}
  \bibinfo{author}{\bibfnamefont{E.}~\bibnamefont{Tosatti}},
  \bibinfo{journal}{Surf. Sci.} \textbf{\bibinfo{volume}{454}},
  \bibinfo{pages}{172} (\bibinfo{year}{2000}).

\bibitem[{\citenamefont{P\'{e}rez et~al.}(2001)\citenamefont{P\'{e}rez, Ortega,
  and Flores}}]{per01}
\bibinfo{author}{\bibfnamefont{R.}~\bibnamefont{P\'{e}rez}},
  \bibinfo{author}{\bibfnamefont{J.}~\bibnamefont{Ortega}}, \bibnamefont{and}
  \bibinfo{author}{\bibfnamefont{F.}~\bibnamefont{Flores}},
  \bibinfo{journal}{Phys. Rev. Lett.} \textbf{\bibinfo{volume}{86}},
  \bibinfo{pages}{4891} (\bibinfo{year}{2001}).

\bibitem[{\citenamefont{Ballabio et~al.}(2002)\citenamefont{Ballabio, Profeta,
  {de~Gironcoli}, Scandolo, Santoro, and Tosatti}}]{bal02}
\bibinfo{author}{\bibfnamefont{G.}~\bibnamefont{Ballabio}},
  \bibinfo{author}{\bibfnamefont{G.}~\bibnamefont{Profeta}},
  \bibinfo{author}{\bibfnamefont{S.}~\bibnamefont{{de~Gironcoli}}},
  \bibinfo{author}{\bibfnamefont{S.}~\bibnamefont{Scandolo}},
  \bibinfo{author}{\bibfnamefont{G.~E.} \bibnamefont{Santoro}},
  \bibnamefont{and} \bibinfo{author}{\bibfnamefont{E.}~\bibnamefont{Tosatti}},
  \bibinfo{journal}{Phys. Rev. Lett.} \textbf{\bibinfo{volume}{89}},
  \bibinfo{pages}{126803} (\bibinfo{year}{2002}).

\bibitem[{\citenamefont{Gori et~al.}(2009)\citenamefont{Gori, Ronci, Colonna,
  Cricenti, Pulci, and LeLay}}]{gor09}
\bibinfo{author}{\bibfnamefont{P.}~\bibnamefont{Gori}},
  \bibinfo{author}{\bibfnamefont{F.}~\bibnamefont{Ronci}},
  \bibinfo{author}{\bibfnamefont{S.}~\bibnamefont{Colonna}},
  \bibinfo{author}{\bibfnamefont{A.}~\bibnamefont{Cricenti}},
  \bibinfo{author}{\bibfnamefont{O.}~\bibnamefont{Pulci}}, \bibnamefont{and}
  \bibinfo{author}{\bibfnamefont{G.}~\bibnamefont{LeLay}},
  \bibinfo{journal}{Europhys. Lett.} \textbf{\bibinfo{volume}{85}},
  \bibinfo{pages}{66001} (\bibinfo{year}{2009}).

\bibitem[{\citenamefont{Profeta and Tosatti}(2007)}]{pro07}
\bibinfo{author}{\bibfnamefont{G.}~\bibnamefont{Profeta}} \bibnamefont{and}
  \bibinfo{author}{\bibfnamefont{E.}~\bibnamefont{Tosatti}},
  \bibinfo{journal}{Phys. Rev. Lett.} \textbf{\bibinfo{volume}{98}},
  \bibinfo{pages}{086401} (\bibinfo{year}{2007}).

\bibitem[{\citenamefont{Anisimov et~al.}(1991)\citenamefont{Anisimov, Zaanen,
  and Andersen}}]{ani_ldau}
\bibinfo{author}{\bibfnamefont{V.~I.} \bibnamefont{Anisimov}},
  \bibinfo{author}{\bibfnamefont{J.}~\bibnamefont{Zaanen}}, \bibnamefont{and}
  \bibinfo{author}{\bibfnamefont{O.~K.} \bibnamefont{Andersen}},
  \bibinfo{journal}{Phys. Rev. B} \textbf{\bibinfo{volume}{44}},
  \bibinfo{pages}{943} (\bibinfo{year}{1991}).

\bibitem[{\citenamefont{Anisimov et~al.}(1997)\citenamefont{Anisimov,
  Poteryaev, Korotin, Anokhin, and Kotliar}}]{ani97}
\bibinfo{author}{\bibfnamefont{V.~I.} \bibnamefont{Anisimov}},
  \bibinfo{author}{\bibfnamefont{A.~I.} \bibnamefont{Poteryaev}},
  \bibinfo{author}{\bibfnamefont{M.~A.} \bibnamefont{Korotin}},
  \bibinfo{author}{\bibfnamefont{A.~O.} \bibnamefont{Anokhin}},
  \bibnamefont{and} \bibinfo{author}{\bibfnamefont{G.}~\bibnamefont{Kotliar}},
  \bibinfo{journal}{J. Phys.: Condens. Matter} \textbf{\bibinfo{volume}{9}},
  \bibinfo{pages}{7359} (\bibinfo{year}{1997}).

\bibitem[{\citenamefont{Lichtenstein and Katsnelson}(1998)}]{lic98}
\bibinfo{author}{\bibfnamefont{A.~I.} \bibnamefont{Lichtenstein}}
  \bibnamefont{and} \bibinfo{author}{\bibfnamefont{M.~I.}
  \bibnamefont{Katsnelson}}, \bibinfo{journal}{Phys. Rev. B}
  \textbf{\bibinfo{volume}{57}}, \bibinfo{pages}{6884} (\bibinfo{year}{1998}).

\bibitem[{\citenamefont{Georges and Kotliar}(1992)}]{geo92}
\bibinfo{author}{\bibfnamefont{A.}~\bibnamefont{Georges}} \bibnamefont{and}
  \bibinfo{author}{\bibfnamefont{G.}~\bibnamefont{Kotliar}},
  \bibinfo{journal}{Phys. Rev. B} \textbf{\bibinfo{volume}{45}},
  \bibinfo{pages}{6479} (\bibinfo{year}{1992}).

\bibitem[{\citenamefont{Metzner and Vollhardt}(1989)}]{met89}
\bibinfo{author}{\bibfnamefont{W.}~\bibnamefont{Metzner}} \bibnamefont{and}
  \bibinfo{author}{\bibfnamefont{D.}~\bibnamefont{Vollhardt}},
  \bibinfo{journal}{Phys. Rev. Lett.} \textbf{\bibinfo{volume}{62}},
  \bibinfo{pages}{324} (\bibinfo{year}{1989}).

\bibitem[{\citenamefont{B\"{u}nemann et~al.}(2003)\citenamefont{B\"{u}nemann,
  Gebhard, Ohm, Umst\"atter, Weiser, Weber, Claessen, Ehm, Harasawa, Kakizaki
  et~al.}}]{bue03}
\bibinfo{author}{\bibfnamefont{J.}~\bibnamefont{B\"{u}nemann}},
  \bibinfo{author}{\bibfnamefont{F.}~\bibnamefont{Gebhard}},
  \bibinfo{author}{\bibfnamefont{T.}~\bibnamefont{Ohm}},
  \bibinfo{author}{\bibfnamefont{R.}~\bibnamefont{Umst\"atter}},
  \bibinfo{author}{\bibfnamefont{S.}~\bibnamefont{Weiser}},
  \bibinfo{author}{\bibfnamefont{W.}~\bibnamefont{Weber}},
  \bibinfo{author}{\bibfnamefont{R.}~\bibnamefont{Claessen}},
  \bibinfo{author}{\bibfnamefont{D.}~\bibnamefont{Ehm}},
  \bibinfo{author}{\bibfnamefont{A.}~\bibnamefont{Harasawa}},
  \bibinfo{author}{\bibfnamefont{A.}~\bibnamefont{Kakizaki}},
  \bibnamefont{et~al.}, \bibinfo{journal}{Europhys. Lett.}
  \textbf{\bibinfo{volume}{61}}, \bibinfo{pages}{667} (\bibinfo{year}{2003}).

\bibitem[{\citenamefont{Deng et~al.}(2008)\citenamefont{Deng, Dai, and
  Fang}}]{den08}
\bibinfo{author}{\bibfnamefont{X.~Y.} \bibnamefont{Deng}},
  \bibinfo{author}{\bibfnamefont{X.}~\bibnamefont{Dai}}, \bibnamefont{and}
  \bibinfo{author}{\bibfnamefont{Z.}~\bibnamefont{Fang}},
  \bibinfo{journal}{Europhys. Lett.} \textbf{\bibinfo{volume}{83}},
  \bibinfo{pages}{37008} (\bibinfo{year}{2008}).

\bibitem[{\citenamefont{Lechermann}(2009)}]{lec09}
\bibinfo{author}{\bibfnamefont{F.}~\bibnamefont{Lechermann}},
  \bibinfo{journal}{Phys. Rev. Lett.} \textbf{\bibinfo{volume}{102}},
  \bibinfo{pages}{046403} (\bibinfo{year}{2009}).

\bibitem[{\citenamefont{B\"{u}nemann et~al.}(1998)\citenamefont{B\"{u}nemann,
  Weber, and Gebhard}}]{bue98}
\bibinfo{author}{\bibfnamefont{J.}~\bibnamefont{B\"{u}nemann}},
  \bibinfo{author}{\bibfnamefont{W.}~\bibnamefont{Weber}}, \bibnamefont{and}
  \bibinfo{author}{\bibfnamefont{F.}~\bibnamefont{Gebhard}},
  \bibinfo{journal}{Phys. Rev. B} \textbf{\bibinfo{volume}{57}},
  \bibinfo{pages}{6896} (\bibinfo{year}{1998}).

\bibitem[{\citenamefont{Lechermann et~al.}(2007)\citenamefont{Lechermann,
  Georges, Kotliar, and Parcollet}}]{lec07}
\bibinfo{author}{\bibfnamefont{F.}~\bibnamefont{Lechermann}},
  \bibinfo{author}{\bibfnamefont{A.}~\bibnamefont{Georges}},
  \bibinfo{author}{\bibfnamefont{G.}~\bibnamefont{Kotliar}}, \bibnamefont{and}
  \bibinfo{author}{\bibfnamefont{O.}~\bibnamefont{Parcollet}},
  \bibinfo{journal}{Phys. Rev. B} \textbf{\bibinfo{volume}{76}},
  \bibinfo{pages}{155102} (\bibinfo{year}{2007}).

\bibitem[{\citenamefont{Fabrizio}(2007)}]{fab07}
\bibinfo{author}{\bibfnamefont{M.}~\bibnamefont{Fabrizio}},
  \bibinfo{journal}{Phys. Rev. B} \textbf{\bibinfo{volume}{76}},
  \bibinfo{pages}{165110} (\bibinfo{year}{2007}).

\bibitem[{\citenamefont{Meyer et~al.}(unpublished)\citenamefont{Meyer,
  Els\"{a}sser, Lechermann, and F\"{a}hnle}}]{mbpp_code}
\bibinfo{author}{\bibfnamefont{B.}~\bibnamefont{Meyer}},
  \bibinfo{author}{\bibfnamefont{C.}~\bibnamefont{Els\"{a}sser}},
  \bibinfo{author}{\bibfnamefont{F.}~\bibnamefont{Lechermann}},
  \bibnamefont{and}
  \bibinfo{author}{\bibfnamefont{M.}~\bibnamefont{F\"{a}hnle}},
  \emph{\bibinfo{title}{FORTRAN 90 Program for Mixed-Basis-Pseudopotential
  Calculations for Crystals}}, \bibinfo{organization}{Max-Planck-Institut
  f\"{u}r Metallforschung, Stuttgart} (\bibinfo{year}{unpublished}).

\bibitem[{\citenamefont{Louie et~al.}(1979)\citenamefont{Louie, Ho, and
  Cohen}}]{lou79}
\bibinfo{author}{\bibfnamefont{S.~G.} \bibnamefont{Louie}},
  \bibinfo{author}{\bibfnamefont{K.~M.} \bibnamefont{Ho}}, \bibnamefont{and}
  \bibinfo{author}{\bibfnamefont{M.~L.} \bibnamefont{Cohen}},
  \bibinfo{journal}{Phys. Rev. B} \textbf{\bibinfo{volume}{19}},
  \bibinfo{pages}{1774} (\bibinfo{year}{1979}).

\bibitem[{\citenamefont{Vanderbilt}(1985)}]{van85}
\bibinfo{author}{\bibfnamefont{D.}~\bibnamefont{Vanderbilt}},
  \bibinfo{journal}{Phys. Rev. B} \textbf{\bibinfo{volume}{32}},
  \bibinfo{pages}{8412} (\bibinfo{year}{1985}).

\bibitem[{\citenamefont{Bl\"ochl}(1994)}]{blo94}
\bibinfo{author}{\bibfnamefont{P.~E.} \bibnamefont{Bl\"ochl}},
  \bibinfo{journal}{Phys. Rev. B} \textbf{\bibinfo{volume}{50}},
  \bibinfo{pages}{17953} (\bibinfo{year}{1994}).

\bibitem[{\citenamefont{Kresse and Hafner}(1994)}]{kre94}
\bibinfo{author}{\bibfnamefont{G.}~\bibnamefont{Kresse}} \bibnamefont{and}
  \bibinfo{author}{\bibfnamefont{J.}~\bibnamefont{Hafner}},
  \bibinfo{journal}{J. Phys.: Condens. Matter} \textbf{\bibinfo{volume}{6}},
  \bibinfo{pages}{8245} (\bibinfo{year}{1994}).

\bibitem[{\citenamefont{Hirsch and Fye}(1986)}]{hir86}
\bibinfo{author}{\bibfnamefont{J.~E.} \bibnamefont{Hirsch}} \bibnamefont{and}
  \bibinfo{author}{\bibfnamefont{R.~M.} \bibnamefont{Fye}},
  \bibinfo{journal}{Phys. Rev. Lett.} \textbf{\bibinfo{volume}{56}},
  \bibinfo{pages}{2521} (\bibinfo{year}{1986}).

\bibitem[{\citenamefont{Li et~al.}(1989)\citenamefont{Li, W\"olfle, and
  Hirschfeld}}]{li89}
\bibinfo{author}{\bibfnamefont{T.}~\bibnamefont{Li}},
  \bibinfo{author}{\bibfnamefont{P.}~\bibnamefont{W\"olfle}}, \bibnamefont{and}
  \bibinfo{author}{\bibfnamefont{P.~J.} \bibnamefont{Hirschfeld}},
  \bibinfo{journal}{Phys. Rev. B} \textbf{\bibinfo{volume}{40}},
  \bibinfo{pages}{6817} (\bibinfo{year}{1989}).

\bibitem[{\citenamefont{B\"{u}nemann and Gebhard}(2007)}]{bue07}
\bibinfo{author}{\bibfnamefont{J.}~\bibnamefont{B\"{u}nemann}}
  \bibnamefont{and} \bibinfo{author}{\bibfnamefont{F.}~\bibnamefont{Gebhard}},
  \bibinfo{journal}{Phys. Rev. B} \textbf{\bibinfo{volume}{76}},
  \bibinfo{pages}{193104} (\bibinfo{year}{2007}).

\bibitem[{\citenamefont{Lechermann et~al.}(2006)\citenamefont{Lechermann,
  Georges, Poteryaev, Biermann, Posternak, Yamasaki, and Andersen}}]{lec06}
\bibinfo{author}{\bibfnamefont{F.}~\bibnamefont{Lechermann}},
  \bibinfo{author}{\bibfnamefont{A.}~\bibnamefont{Georges}},
  \bibinfo{author}{\bibfnamefont{A.}~\bibnamefont{Poteryaev}},
  \bibinfo{author}{\bibfnamefont{S.}~\bibnamefont{Biermann}},
  \bibinfo{author}{\bibfnamefont{M.}~\bibnamefont{Posternak}},
  \bibinfo{author}{\bibfnamefont{A.}~\bibnamefont{Yamasaki}}, \bibnamefont{and}
  \bibinfo{author}{\bibfnamefont{O.~K.} \bibnamefont{Andersen}},
  \bibinfo{journal}{Phys. Rev. B} \textbf{\bibinfo{volume}{74}},
  \bibinfo{pages}{125120} (\bibinfo{year}{2006}).

\bibitem[{\citenamefont{Kotliar et~al.}(2006)\citenamefont{Kotliar, Savrasov,
  Haule, Oudovenko, Parcollet, and Marianetti}}]{kotliar_review}
\bibinfo{author}{\bibfnamefont{G.}~\bibnamefont{Kotliar}},
  \bibinfo{author}{\bibfnamefont{S.~Y.} \bibnamefont{Savrasov}},
  \bibinfo{author}{\bibfnamefont{K.}~\bibnamefont{Haule}},
  \bibinfo{author}{\bibfnamefont{V.~S.} \bibnamefont{Oudovenko}},
  \bibinfo{author}{\bibfnamefont{O.}~\bibnamefont{Parcollet}},
  \bibnamefont{and} \bibinfo{author}{\bibfnamefont{C.~A.}
  \bibnamefont{Marianetti}}, \bibinfo{journal}{Rev. Mod. Phys.}
  \textbf{\bibinfo{volume}{78}}, \bibinfo{pages}{865} (\bibinfo{year}{2006}).

\bibitem[{\citenamefont{Anisimov et~al.}(2005)\citenamefont{Anisimov, Kondakov,
  Kozhevnikov, Nekrasov, Pchelkina, Allen, Mo, Kim, Metcalf, Suga
  et~al.}}]{ani05}
\bibinfo{author}{\bibfnamefont{V.~I.} \bibnamefont{Anisimov}},
  \bibinfo{author}{\bibfnamefont{D.~E.} \bibnamefont{Kondakov}},
  \bibinfo{author}{\bibfnamefont{A.~V.} \bibnamefont{Kozhevnikov}},
  \bibinfo{author}{\bibfnamefont{I.~A.} \bibnamefont{Nekrasov}},
  \bibinfo{author}{\bibfnamefont{Z.~V.} \bibnamefont{Pchelkina}},
  \bibinfo{author}{\bibfnamefont{J.~W.} \bibnamefont{Allen}},
  \bibinfo{author}{\bibfnamefont{S.-K.} \bibnamefont{Mo}},
  \bibinfo{author}{\bibfnamefont{H.-D.} \bibnamefont{Kim}},
  \bibinfo{author}{\bibfnamefont{P.}~\bibnamefont{Metcalf}},
  \bibinfo{author}{\bibfnamefont{S.}~\bibnamefont{Suga}}, \bibnamefont{et~al.},
  \bibinfo{journal}{Phys. Rev. B} \textbf{\bibinfo{volume}{71}},
  \bibinfo{pages}{125119} (\bibinfo{year}{2005}).

\bibitem[{\citenamefont{Marzari and Vanderbilt}(1997)}]{mar97}
\bibinfo{author}{\bibfnamefont{N.}~\bibnamefont{Marzari}} \bibnamefont{and}
  \bibinfo{author}{\bibfnamefont{D.}~\bibnamefont{Vanderbilt}},
  \bibinfo{journal}{Phys. Rev. B} \textbf{\bibinfo{volume}{56}},
  \bibinfo{pages}{12847} (\bibinfo{year}{1997}).

\bibitem[{\citenamefont{Souza et~al.}(2001)\citenamefont{Souza, Marzari, and
  Vanderbilt}}]{sou01}
\bibinfo{author}{\bibfnamefont{I.}~\bibnamefont{Souza}},
  \bibinfo{author}{\bibfnamefont{N.}~\bibnamefont{Marzari}}, \bibnamefont{and}
  \bibinfo{author}{\bibfnamefont{D.}~\bibnamefont{Vanderbilt}},
  \bibinfo{journal}{Phys. Rev. B} \textbf{\bibinfo{volume}{65}},
  \bibinfo{pages}{035109} (\bibinfo{year}{2001}).

\bibitem[{\citenamefont{Perdew et~al.}(1996)\citenamefont{Perdew, Burke, and
  Ernzerhof}}]{per96}
\bibinfo{author}{\bibfnamefont{J.~P.} \bibnamefont{Perdew}},
  \bibinfo{author}{\bibfnamefont{K.}~\bibnamefont{Burke}}, \bibnamefont{and}
  \bibinfo{author}{\bibfnamefont{M.}~\bibnamefont{Ernzerhof}},
  \bibinfo{journal}{Phys. Rev. Lett.} \textbf{\bibinfo{volume}{77}},
  \bibinfo{pages}{3865} (\bibinfo{year}{1996}).

\bibitem[{\citenamefont{Bachelet and Christensen}(1985)}]{bac85}
\bibinfo{author}{\bibfnamefont{G.~B.} \bibnamefont{Bachelet}} \bibnamefont{and}
  \bibinfo{author}{\bibfnamefont{N.~E.} \bibnamefont{Christensen}},
  \bibinfo{journal}{Phys. Rev. B} \textbf{\bibinfo{volume}{31}},
  \bibinfo{pages}{879} (\bibinfo{year}{1985}).

\bibitem[{\citenamefont{Gl\"otzel et~al.}(1980)\citenamefont{Gl\"otzel, Segall,
  and Andersen}}]{glo80}
\bibinfo{author}{\bibfnamefont{D.}~\bibnamefont{Gl\"otzel}},
  \bibinfo{author}{\bibfnamefont{B.}~\bibnamefont{Segall}}, \bibnamefont{and}
  \bibinfo{author}{\bibfnamefont{O.~K.} \bibnamefont{Andersen}},
  \bibinfo{journal}{Solid State Comm.} \textbf{\bibinfo{volume}{36}},
  \bibinfo{pages}{403} (\bibinfo{year}{1980}).

\bibitem[{\citenamefont{Mielke}(1991)}]{mie91}
\bibinfo{author}{\bibfnamefont{A.}~\bibnamefont{Mielke}},
  \bibinfo{journal}{J.~Phys. A} \textbf{\bibinfo{volume}{24}},
  \bibinfo{pages}{L73} (\bibinfo{year}{1991}).

\bibitem[{\citenamefont{Tasaki}(1992)}]{tas92}
\bibinfo{author}{\bibfnamefont{H.}~\bibnamefont{Tasaki}},
  \bibinfo{journal}{Phys. Rev. Lett.} \textbf{\bibinfo{volume}{69}},
  \bibinfo{pages}{1608} (\bibinfo{year}{1992}).

\bibitem[{\citenamefont{Lichtenstein et~al.}(2002)\citenamefont{Lichtenstein,
  Katsnelson, and Kotliar}}]{lic02}
\bibinfo{author}{\bibfnamefont{A.}~\bibnamefont{Lichtenstein}},
  \bibinfo{author}{\bibfnamefont{M.}~\bibnamefont{Katsnelson}},
  \bibnamefont{and} \bibinfo{author}{\bibfnamefont{G.}~\bibnamefont{Kotliar}},
  \bibinfo{journal}{cond-mat/0211076}  (\bibinfo{year}{2002}).

\bibitem[{\citenamefont{Biroli et~al.}(2004)\citenamefont{Biroli, Parcollet,
  and Kotliar}}]{bir04}
\bibinfo{author}{\bibfnamefont{G.}~\bibnamefont{Biroli}},
  \bibinfo{author}{\bibfnamefont{O.}~\bibnamefont{Parcollet}},
  \bibnamefont{and} \bibinfo{author}{\bibfnamefont{G.}~\bibnamefont{Kotliar}},
  \bibinfo{journal}{Phys. Rev. B} \textbf{\bibinfo{volume}{69}},
  \bibinfo{pages}{205108} (\bibinfo{year}{2004}).

\bibitem[{\citenamefont{Maier et~al.}(2005)\citenamefont{Maier, Jarrell,
  Pruschke, and Hettler}}]{mai05}
\bibinfo{author}{\bibfnamefont{T.}~\bibnamefont{Maier}},
  \bibinfo{author}{\bibfnamefont{M.}~\bibnamefont{Jarrell}},
  \bibinfo{author}{\bibfnamefont{T.}~\bibnamefont{Pruschke}}, \bibnamefont{and}
  \bibinfo{author}{\bibfnamefont{M.~H.} \bibnamefont{Hettler}},
  \bibinfo{journal}{Rev. Mod. Phys.} \textbf{\bibinfo{volume}{77}},
  \bibinfo{pages}{1027} (\bibinfo{year}{2005}).

\bibitem[{\citenamefont{Tejeda et~al.}(2008)\citenamefont{Tejeda, Cort{\'e}s,
  Lobo-Checa, Didiot, Kierren, Malterre, Michel, and Mascaraque}}]{tej08}
\bibinfo{author}{\bibfnamefont{A.}~\bibnamefont{Tejeda}},
  \bibinfo{author}{\bibfnamefont{R.}~\bibnamefont{Cort{\'e}s}},
  \bibinfo{author}{\bibfnamefont{J.}~\bibnamefont{Lobo-Checa}},
  \bibinfo{author}{\bibfnamefont{C.}~\bibnamefont{Didiot}},
  \bibinfo{author}{\bibfnamefont{B.}~\bibnamefont{Kierren}},
  \bibinfo{author}{\bibfnamefont{D.}~\bibnamefont{Malterre}},
  \bibinfo{author}{\bibfnamefont{E.~G.} \bibnamefont{Michel}},
  \bibnamefont{and}
  \bibinfo{author}{\bibfnamefont{A.}~\bibnamefont{Mascaraque}},
  \bibinfo{journal}{Phys. Rev. Lett.} \textbf{\bibinfo{volume}{100}},
  \bibinfo{pages}{026103} (\bibinfo{year}{2008}).

\end{thebibliography}

\end{document}